\documentclass[epjST]{svjour}
\usepackage{amsmath}
\usepackage{subfigure}
\usepackage{multirow}
\usepackage{graphics}
\usepackage{graphicx}
\usepackage{epsfig}

 \DeclareMathOperator{\erf}{erf} 
\DeclareMathOperator{\expon}{exp}

\begin{document}
\title{Effects of immersed moonlets in the ring arc particles of Saturn }
\author{G. Madeira\inst{1}\fnmsep\thanks{\email{gustavo.o.madeira@unesp.br
}} \and S.M. Giuliatti Winter\inst{1}}
\institute{Grupo de Din\^amica Orbital \& Planetologia - Univ. of the State of S\~ao Paulo - UNESP - Guaratinguet\'a - Brazil}
\abstract{
Ring arcs  are the result of particles in corotation resonances with nearby satellites. Arcs are present in Saturn and Neptune systems,  in Saturn they are also associated with small satellites immersed on them. The satellite Aegaeon is immersed in the G~ring arc, and the satellites Anthe and Methone are embedded in arcs named after them. Since most of the population of the arcs is formed by $\mu$m-sized particles the dissipative effects, such as the plasma drag and the solar radiation  force, decrease the lifetime of the arcs. We analysed the effects of the immersed satellites  on these arcs by computing the mass production rate and the  perturbation caused by them in the arc particles.  By comparing the lifetime of the particles and the mass production rate we concluded that Aegaeon, Anthe and Methone did not act as sources for their arcs. We took a step further by analysing a hypothetical scenario formed by an immersed  moonlet  of different sizes.  As a result we found that regardless the size of the hypothetical moonlet (from about 0.10 km to 4.0 km) these moonlets will not act as a source. These arcs are temporary structures and they will disappear in a very short period of time.} 
\maketitle

\section{Introduction}
\label{intro}

Corotation resonances  between ring  particles and  nearby satellites  are responsible for the formation  of arcs in planetary rings. The first arcs to be discovered were the arcs of  Neptune Adams ring. These arcs, named Courage, Libert\'e, Egalit\'e and Fraternit\'e, are the densest parts of the Adams ring \cite{Po91}. Recent data  have shown changes in the intensity of these arcs \cite{Pa05} raising the question that these structures can be transient.

The arcs and small satellites of Saturn were discovered by Cassini images: the G~ring arc located between the orbits of the satellites Janus/Epimetheus and Mimas \cite{He07} and the arcs of Anthe and Methone, located between the satellites Mimas and Enceladus, immersed in the E~ring \cite{He09}.
These arcs are mainly populated by $\mu$m-sized dust particles, althoug Cassini data indicate that these structures may have large bodies.

Small satellites are immersed in these arcs. Aegaeon is a satelite of  500~m in diameter embedded in the G ring arc \cite{He10}. The satellites Anthe \cite{Co08} and Methone \cite{Sp06} are immersed in the arcs named after them with radius of about  500~m and 1450~m \cite{Th13}, respectively.

 References \cite{He10}, \cite{Co08} and \cite{Sp06} have shown that Aegaeon, Methone and Anthe are trapped in a 7:6, 14:15 and 10:11 corotation eccentric resonance (CER) with Mimas, respectively. These arcs are also influenced by nearby Lindblad eccentric resonance (LER). 
 \cite{Mo13} studied the coupling of the corotation and Lindblad resonance through a Hamiltonian formalism, the CoraLin Model, and showed that Aegaeon, Anthe and Methone are close to chaotic regions.

\cite{He10} argued that embedded satellites could produce  dust due to impacts between interplanetary grains onto the surface of these satellites. These dust particles  can replace those particles  lost by ejection and collisions. However, \cite{Ma18} showed that despite the material produced by Aegaeon, the G~ring arc is depleted of $\mu$m-sized particles in less than 40~years due to the dissipative effects of the solar radiation force. Results presented by \cite{Su17} showed that the arcs of Anthe and Methone lose material produced by their satellites in a few decades also due to dissipative forces.

Through a sample of numerical simulations \cite{Su17}  analysed the system formed by Saturn, Enceladus, Mimas, Anthe and Methone and a set of particles ejected from the surfaces of the small satellites Anthe and Methone. These particles ($ < 40~\mu$m in radius) are also under the effects of the solar radiation force, plasma drag and the electromagnetic force. As a result they found that these particles leave the arcs, in the direction of Enceladus, mainly due to the plasma drag perturbations. The lifetime of these particles is less than few decades.

The Adams ring arcs have also been studied by \cite{Re14} who proposed a model to confine the arcs. Their model consists of a sample of coorbital satellites azimuthally confining the  arc particles, while  the satellite Galatea is responsible for radially confining the arc.  \cite{GW19} included in this dynamical system the effects of the solar radiation force and showed that despite the confinement model the particles can become transient and leave the arcs in a very short time. 

The goal of this work is to analyse the orbital evolution of a sample of arc particles under the effects of dissipative forces and embedded hypothetical moonlets of different sizes. The moonlet will be  responsible for feeding the arcs with dust particles produced by collisions onto its surface, and also for removing the arc particles due to collisions or ejections. 

 The paper is divided into six sections. The introduction is given in section~\ref{intro}, while section~\ref{conservative} deals with the behavior of the particles into corotation eccentric resonance. In section~\ref{dissipative} the behavior of the arc particles is analyzed under the effects of dissipative forces, such as the solar radiation force and plasma drag. In section~\ref{immersed} we derived three different regimes that the particles can follow depending on the size of the moonlet. Section~\ref{fate} analysis the fate of the arcs population by comparing the lifetime of the arcs (with and without dissipative forces) with the production of dust by moonlets of different sizes. Our discussion is presented in section~\ref{discussion}.

%
\section{Conservative dynamics}
\label{conservative}

 The $|m+1|:|m|$ corotation eccentric resonance  occurs when the orbit of a particle is closed  in the rotating frame with the resonant frequency and  the particle is azimuthally confined \cite{Si91}. Mathematically 
\begin{equation}
(m+1)~n_{sat} - m~n - \dot\varpi_{sat} = 0 \label{ressos}
\end{equation}
where $n$ and $n_{sat}$ are the angular frequencies of the particle and the satellite, respectively, and 
$\dot\varpi_{sat}$ is the derivative of the longitude of the  pericentre of the satellite.  For external resonances $m$ is taken as a negative integer and  positive for internal resonances \cite{Si91}.

In this work we analysed the particles located in the arcs of  Aegaeon (G~ring arc), Methone and Anthe  which are in 7:6, 14:15 and 10:11 CER with the satellite Mimas.  The resonant angles are \cite{He10,Sp06,Co08}:
\begin{equation}
\phi_{7:6} =   7\lambda_M - 6\lambda - \varpi_M
\end{equation}
\begin{equation}
\phi_{14:15} = 15\lambda - 14\lambda_M - \varpi_M
\end{equation}
\begin{equation}
\phi_{10:11} = 11\lambda - 10\lambda_M - \varpi_M
\end{equation}
where $\lambda$ is the mean longitude and  $\varpi$ is the longitude of the pericentre.  The subscript $M$ refers to  Mimas and no subscript is related to the arc particle. 

First of all we numerically simulated a sample of particles under the gravitational effects of Saturn and its gravity coefficients ($J_2$, $J_4$ and $J_6$),  and the satellite Mimas. The adopted parameters of Mimas are semi-major axis $a_M = 185539$~km, eccentricity $e_M=0.0196$ and inclination $I_M= 1.574^{\circ}$, obtained from JPL-Horizons System  (2451545.5 JD); the angular parameters were assumed equal to zero. Table~\ref{initial} presents Saturn physical parameters: radius ($R_S$ in km) and mass ($M_M$ in kg) \cite{Th13}, and the gravity coefficients ($J_2$, $J_4$ and $J_6$) derived from \cite{He10}, and the mass of Mimas in kg (from JPL-Horizons System).

\begin{table}
\centering
\caption{Physical parameters of Saturn and the mass of Mimas.}
\label{initial}
\begin{tabular}{ll}
\hline\noalign{\smallskip}
\hline\noalign{\smallskip}
$J_2$ ($\times 10^{-6}$) & $16290.544$ \\
$J_4$ ($\times 10^{-6}$) & $-936.700$ \\
$J_6$ ($\times 10^{-6}$) & $86.623$\\
$R$~(km) & $60330$ \\
$M_S$ ($\times 10^{26}$~kg)& $5.687$ \\
$M_M$ ($\times 10^{19}$~kg) & $3.754$ \\
\hline\noalign{\smallskip}
\end{tabular}
\end{table}

 The angular frequency and semimajor axis are related by the keplerian third law ($GM_S=n^2a^3$, where $G=6.674184\times 10^{-11}~m^{3}kg^{-1}s^{-2}$) and the semimajor axis  resonant ($a_c$) can be determinated  numerically by the Newton-Raphson iterative method \cite{Pr92}: 
\begin{equation}
a_{n+1}=a_n-\frac{\dot{\phi}}{\frac{d\dot{\phi}(a_n)}{da}} \label{raphson} 
\end{equation}
 where $a_n$ and $a_{n+1}$ are the semimajor axis found after the $n_{\rm th}$ and $n_{\rm +1-th}$ iteration, respectively. We assumed an error of $10^{-3}$ km and the initial condition of iteration as:
\begin{equation}
a_0=\left(\frac{m}{m+1}\right)^{\frac{2}{3}}a_M
\end{equation}

 The width of the resonance ($W_c$) is given by \cite{Mu99}:
\begin{equation}
W_c=8a_c\sqrt{\frac{a_M|R^{(c)}|}{3GM_S}}    \label{wc}
\end{equation}

 The term $R^{(c)}$ is the term of the disturbing function associated with the CER \cite{El00}:
\begin{align}
R^{(c)}=\frac{GM_Me_M}{2a_M}\cos{\phi}\left[2m+1+a_c\frac{d}{da_c}\right]b_{1/2}^{(m)} \label{perc}
\end{align}
 where $b_{1/2}^{(m)}$ is the Laplace coefficient.

 The semi-major axis resonant (Equation~\ref{raphson}) and the width of the resonance (Equation~\ref{wc}) are presented in Table~\ref{resonances}.

 The particles were assumed to be in almost circular orbits with $e=10^{- 5}$ orbiting in the equatorial plane of Saturn. A sample of particles were uniformly distributed in the range  $[a_c - W_c/2$, $a_c + W_c/2]$ and mean longitude $[0^\circ$, $360^\circ]$, each 1km and $1^\circ$.  The time span of the numerical simulation is 100~years. We used the Mercury integrator \cite{Ch99} with the Burlish-St\"oer algorithm. The algorithm described in \cite{Re06} was used to convert the state vector into the geometric elements in order to remove the short variations caused by the gravity coefficients. 

\begin{table}
\centering
\caption{Central geometric semimajor axis ($a_c$) of the resonances and the resonance width ($W_c$).} 
\label{resonances}
\begin{tabular}{lcc} 
\hline\noalign{\smallskip}
      & $a_c$ (km) & $W_c$ (km)  \\ \hline\noalign{\smallskip}
$7$:$6$ & $167493.5$ & $64.0$     \\
$14$:$15$ & $194232.9$ & $111.4$     \\
$10$:$11$ & $197655.5$ & $96.3$ \\
\hline\noalign{\smallskip}
\end{tabular}
\end{table}

\begin{figure}
\centering
\subfigure[]{\resizebox{0.95\columnwidth}{!}{\includegraphics{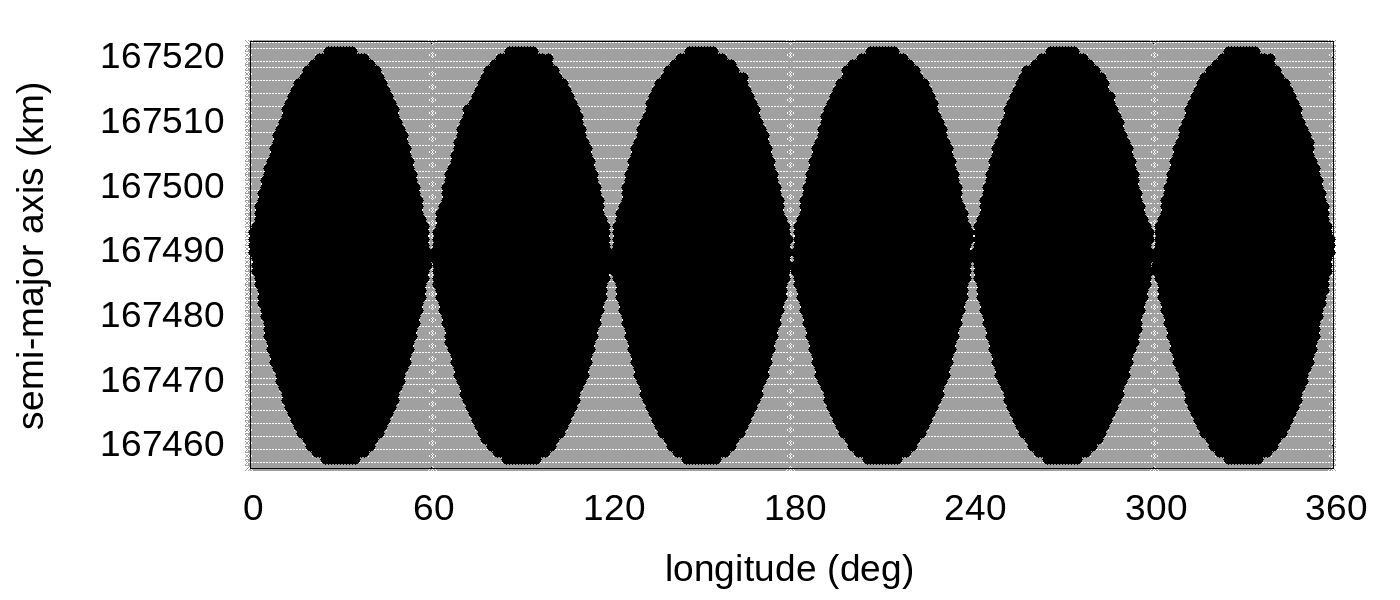}}}
\subfigure[]{\resizebox{0.95\columnwidth}{!}{\includegraphics{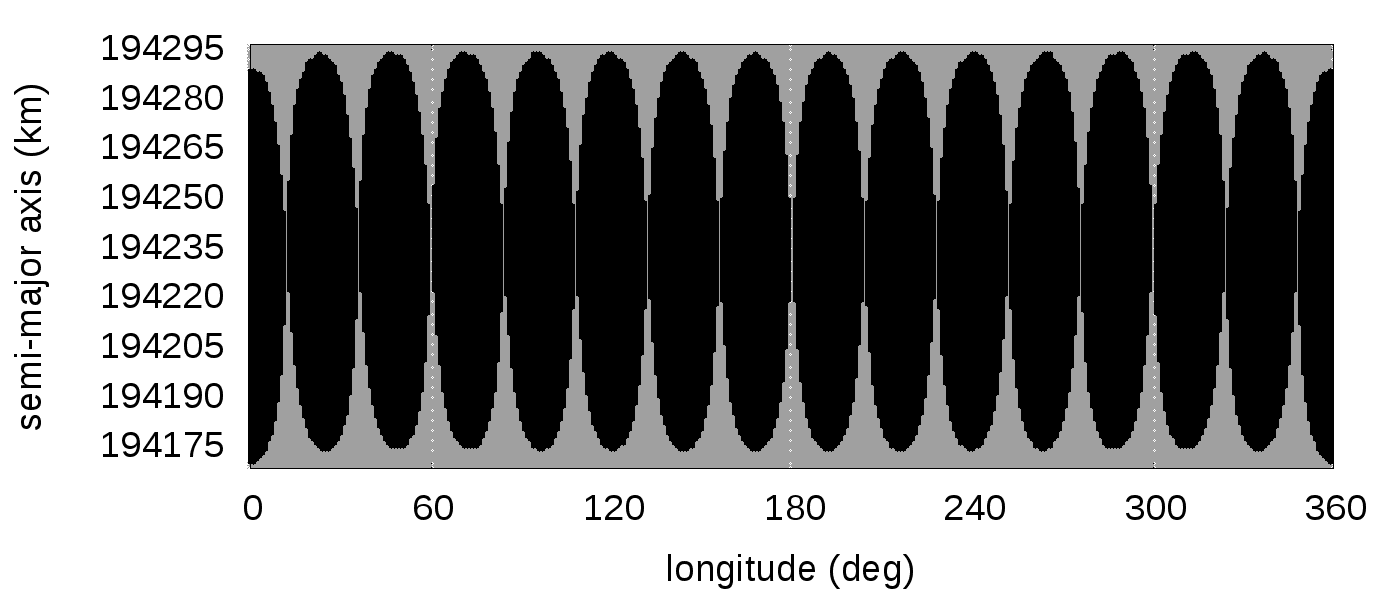}}}
\subfigure[]{\resizebox{0.95\columnwidth}{!}{\includegraphics{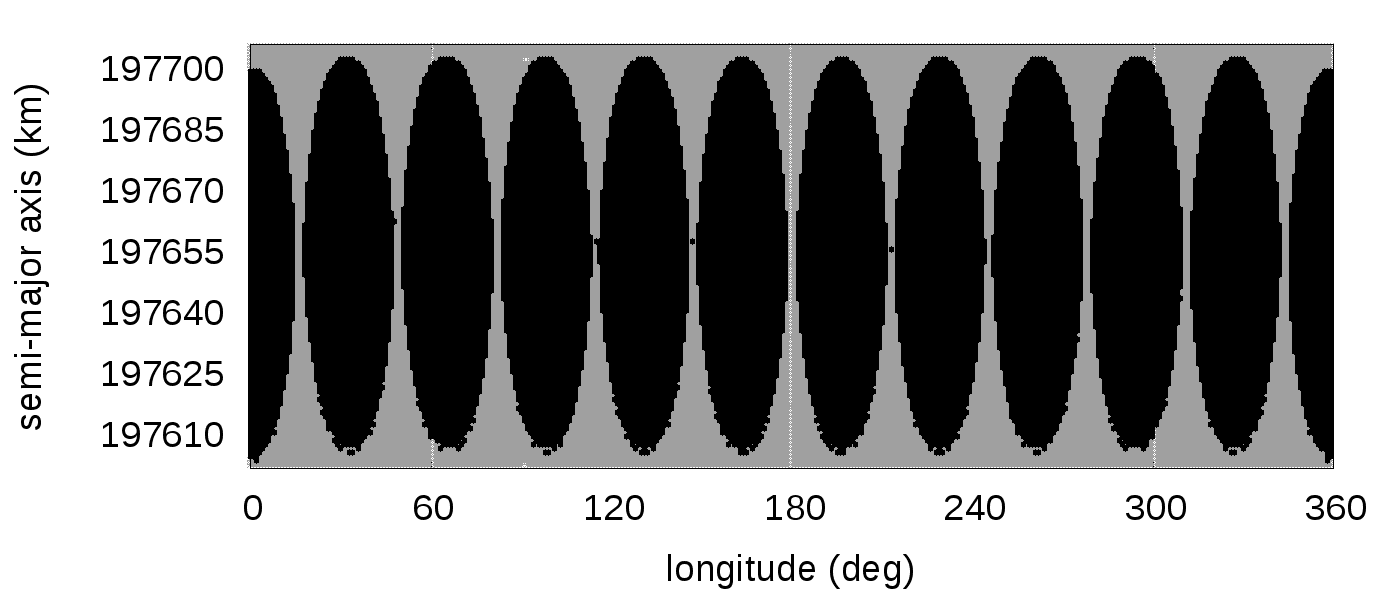}}}
\caption{ In black are shown those particles in resonance with the satellite Mimas: a) 7:6, b) 14:15 and c) 10:11 CER. }
\label{fig:1}       
\end{figure}

 The 7:6, 14:15 and 10:11 CER  produce 6, 15 and 11 equilibrium points, respectively, where the particles move around in the rotating frame with the resonant frequency. The behaviour of the resonant angle of the set of particles has been verified. Figure~\ref{fig:1} shows  the initial semimajor axis and mean longitude of the particles whose resonant angle librates (as shown in black) and in gray are those  particles which resonant angle is circulating.

\begin{figure}
\centering
\resizebox{1.0\columnwidth}{!}{\includegraphics{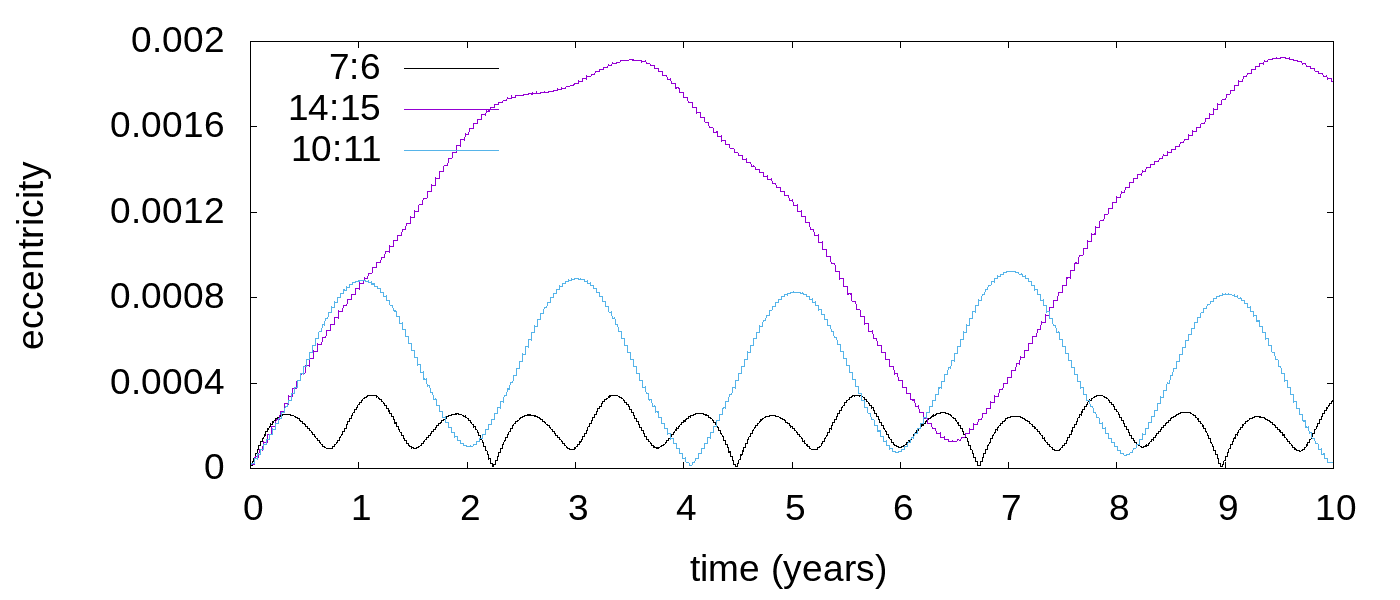}}
\caption{ Time variation of the eccentricity of a representative particle in 7:6, 14:15 and 10:11 CER under the influency of a nearby Lindblad resonance.}
\label{ecc}       
\end{figure}

 Due to the nearby Lindblad resonance, the particles orbits are excited \cite{Mo13} to values of the same order of those observed in the arcs as can be seen in Figure~\ref{ecc} \cite{He09,He10}. This figure shows the time variation of the eccentricity of a representative particle in 7:6, 14:15 and 10:11 CER. We verified that the particles remain azimuthally trapped for a time longer than  $10^5$~years.

\section{Dissipative Forces}
\label{dissipative}

The ring arcs are most composed by $\mu$m-sized  particles  of $1~\mu$m to $10~\mu$m in radius ($r$), as observed by Cassini spacecraft counters \cite{He09,He10}. Solar radiation force and plasma drag are dissipative forces  that can alter the motion of these tiny particles. Therefore we add these forces to the dynamical system in order to analyse the lifetime of the arcs.

The force experienced by a circumplanetary particle due to the solar radiation force (SRF) is given by \cite{Mi84}
\begin{equation}
\vec{F}_{SRF}=\frac{\Phi A}{c} Q_{pr} \left\lbrace \left[1 - \frac{\vec{r}_{sp}}{r_{sp}}\cdot\left( \frac{\vec{V}_{P}}{c} + \frac{\vec{V}}{c} \right) \right] \frac{\vec{r}_{sp}}{r_{sp}}- \frac{\vec{V}_{P} + \vec{V}}{c} \right\rbrace
\label{srforce}
\end{equation}

\noindent where $c$ is the speed of light, $A$ is the cross section of the particles, which are considered to be made of an ideal material ($Q_{PR}=1$) and $\vec{V}$ is the velocity vector of the particle with respect to the planet. The planet is considered to be in a circular heliocentric orbit with position $r_{sp}$ ($r_{sp} = |\vec{r}_{sp}|$)  and velocity $\vec{V}_P$, and $\Phi$ is the solar flux.  Secondary effects such as the Yarkovsky effect, planetary light deflection and shadow were neglected \cite{Ha96}.  

The solar radiation force is composed by two components: the Poynting-Robertson drag and the radiation pressure. The Poynting-Robertson component (the terms that depend on the velocities in equation~\ref{srforce}) provokes a decreasing in the semi-major axis of the particles and the radiation pressure component causes an excitation of the orbits \cite{Sf09}. The  radiation pressure is also responsible for small period variations in the  semi-major axis of the particles. These oscillations are responsible for removing  them from the CER \cite{Ma18}.

Another dissipative force is the plasma drag.
Data obtained by Cassini mission showed that the satellite Enceladus ejects water vapour through geysers near to its south pole. These geysers correspond to the largest source of water group ions (O$^+$, OH$^+$, H$_2$O$^+$ and H$_3$O$^+$) for the saturnian plasma bulk. The O$^+$ ion is the most abundant one \cite{Ma09}.

Therefore, $\mu$m-sized particles located in the  arcs of Methone and Anthe collide with the plasma material, experiencing the plasma drag (PD) \cite{Ba94}
\begin{equation}
\vec{F}_{PD}=\pi n_im_iu_i^2r^2\left[\left(M_i+\frac{1}{2M_i}\right)
\frac{\expon{-M_i^2}}{\sqrt{\pi}} \right. \\
                 \left. +\left(M_i^2+1-\frac{1}{4M_i^2}\right)\erf{M_i}\right]\hat{u}_t    
\label{plasmaforce}
\end{equation}

\noindent  where $n_i$, $u_i$ and $m_i$ are the number density, thermal velocity and mass, respectively, of the ions and $M_i$ is the Mach number.   We assumed the value of $M_i$ relative to O$^+$ to be 1.2 and $n_i = 43.2$~cm$^3$ in the 14:15 region and   $n_i = 44$~cm$^3$ in the 10:11 region \cite{Su17}.  $\hat u_T$ is the unity vector in the tangential direction of the orbit of the particle.

The indirect term generated by the interaction of the plasma material due to Coulomb forces was disregarded, once it is, at least, one order of magnitude weaker \cite{Su17}.

The plasma drag causes an increasing in the semi-major axis of the particles and also changes  in the eccentricities \cite{Ba94}. In the arcs of Anthe and  Methone  this increasing in the semi-major axis is  about $\frac{600}{r(\mu m)}$~km/year \cite{Su17}.  

We numerically simulated one site of each resonance considering particles with sizes of $1~\mu$m and $10~\mu$m  for 1000~years. The particles are under the gravitational effects of Saturn (with the gravity coefficients) and Mimas, and the solar radiation force. Since the effects of the plasma drag on the particles located in the 7:6 CER are two orders of magnitude weaker, we included this  dissipative force only in the  arcs of Anthe and Methone.

\begin{figure}
\centering
\subfigure[]{\resizebox{1.0\columnwidth}{!}{\includegraphics{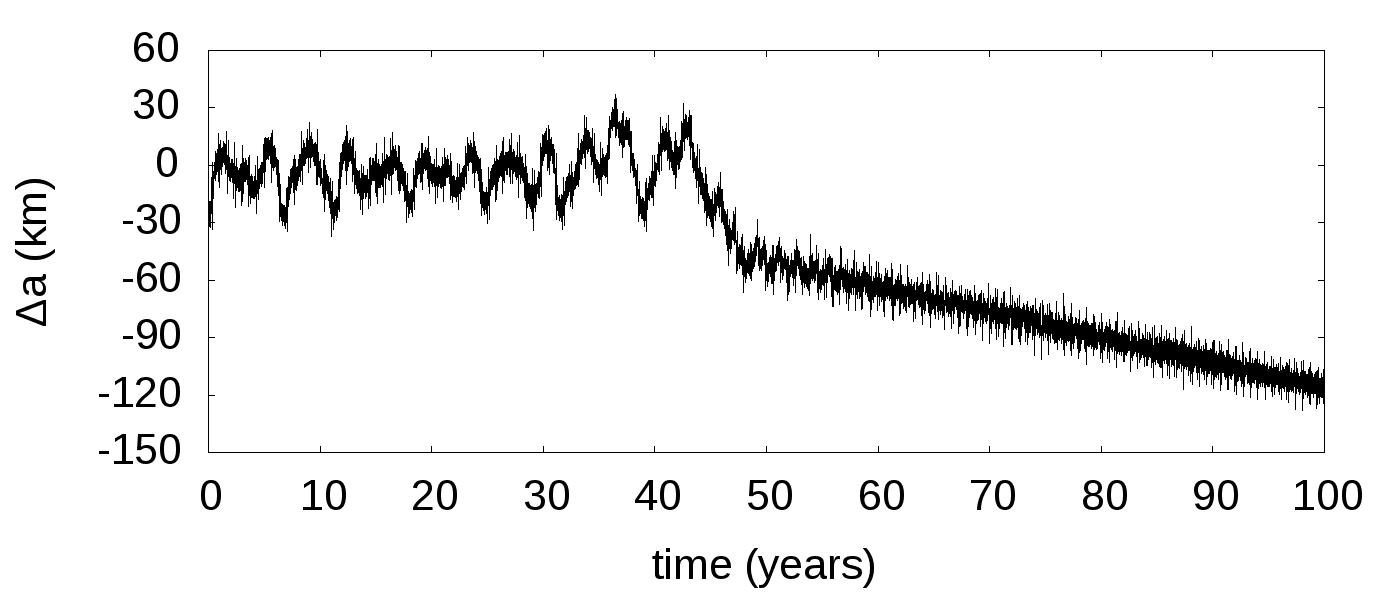}}}
\subfigure[]{\resizebox{1.0\columnwidth}{!}{\includegraphics{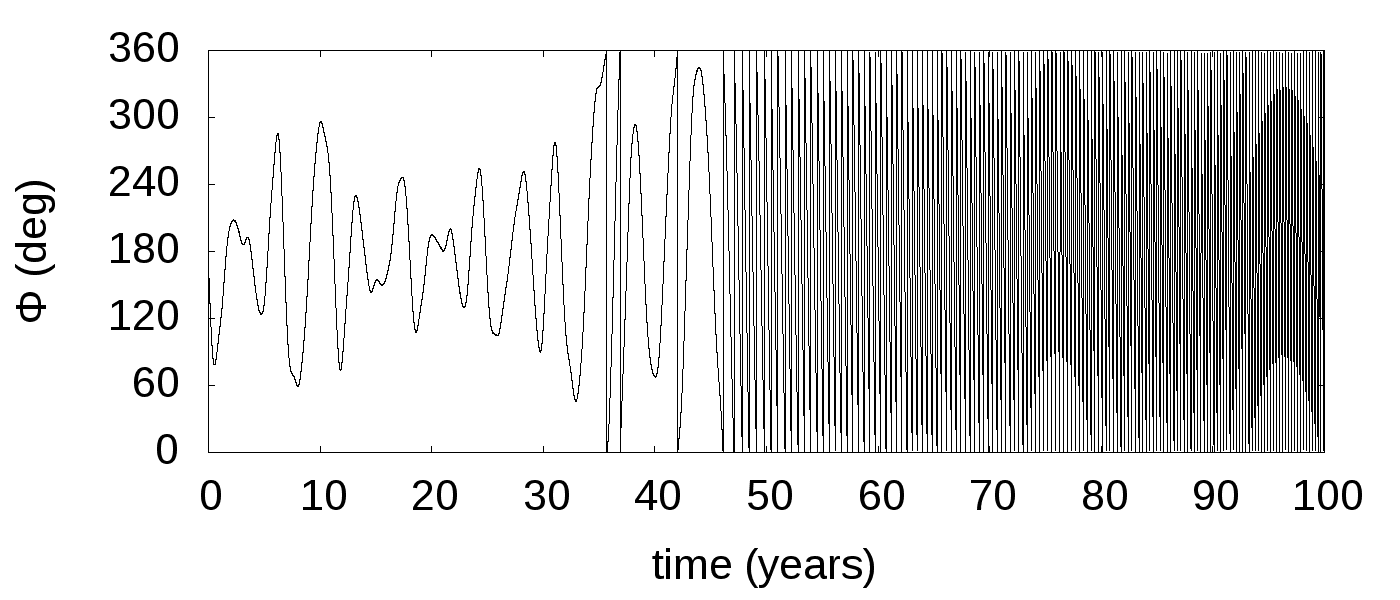}}}
\caption{Time variation of a) $\Delta a$ and b) $\phi$, where $a_c= 167493.5$~km for a $1~\mu$m sized particle initially at $\Delta a = -30$ km and $\Delta \lambda  = 2^\circ$ from the centre of 7:6 CER site.}
\label{fig:2}     
\end{figure}

Figure~\ref{fig:2} shows a) the difference between the semi-major axis of the particle and the centre of the site ($\Delta a$) and b) the  resonant argument ($\phi$) as a function of time. The $1~\mu$m sized particle is initially located at 30~km and $2^\circ$ from the centre of the 7:6 CER site. The particle leaves the resonance after 35 years due to the solar radiation force.  It is trapped  again in resonance and leaves the 7:6 resonance in about 45 years. As can be seen in Figure~\ref{fig:2}a the semimajor axis start decreasing due to the Poynting-Robertson component. Similar results were discussed in \cite{Ma18}.  

\begin{figure}
\centering
\subfigure[]{\resizebox{0.75\columnwidth}{!}{\includegraphics{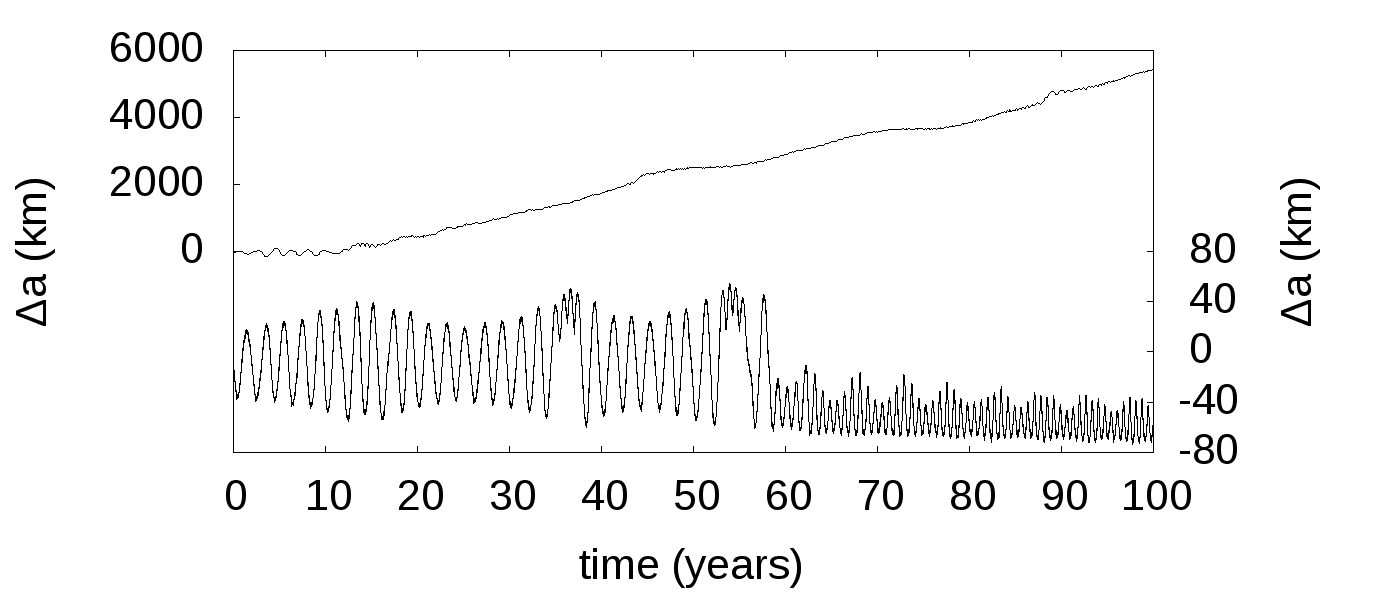}}}
\subfigure[]{\resizebox{0.75\columnwidth}{!}{\includegraphics{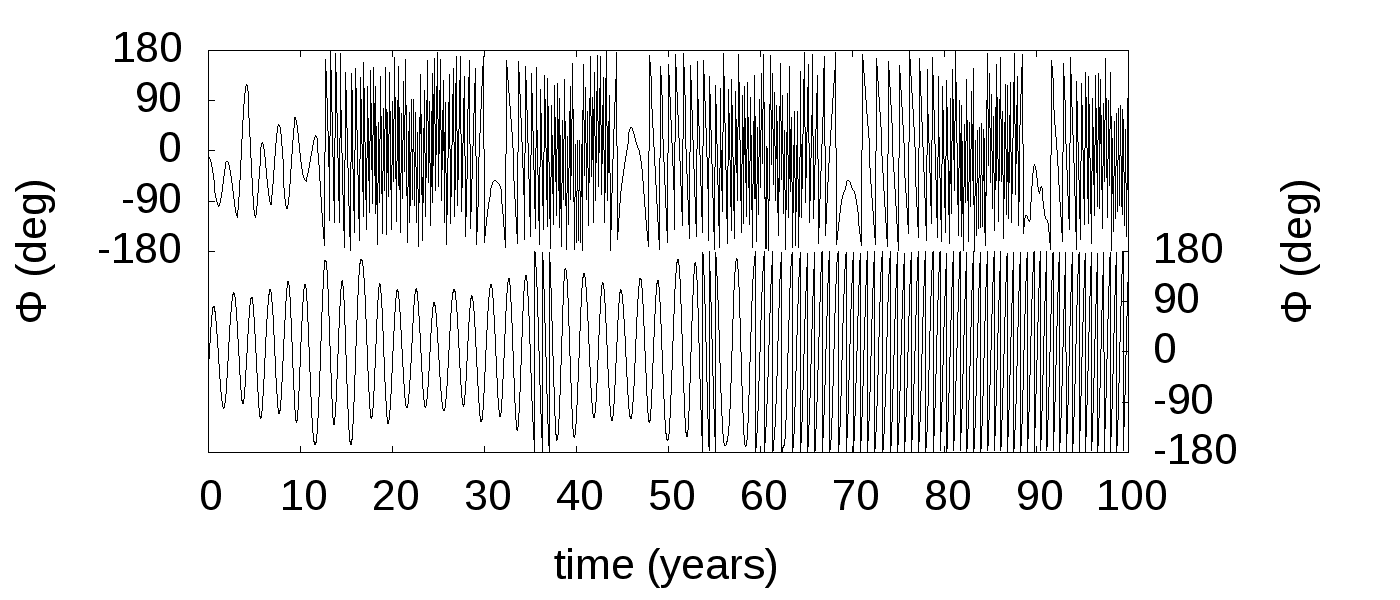}}}
\caption{Time variation of  a) $\Delta a$ and b) $\phi$  for a $10~\mu$m sized particles initially located at $\Delta a = -10$~km and $\Delta \lambda = 0$ from the centre of the 10:11 CER site. In each figure (a and b) there are two curves:  in the top the particle is disturbed by the solar radiation force and the plasma drag while in the bottom the particle is only under the effects of the solar radiation force.}
\label{fig:3}       
\end{figure}

Figure~\ref{fig:3} presents the variation of $\Delta a$ and $\phi$ for a $10~\mu$m sized particle located at the 10:11 CER site.
The particle is initially at $\Delta a = -10$ km and  $\Delta \lambda = 0^\circ$ from the centre of the resonance. In  Figures~\ref{fig:3}a and \ref{fig:3}b there are two curves: in the top the particle is disturbed by the solar radiation force and the plasma drag while in the bottom  the particle is only under the effects of the solar radiation force.
The particle leaves the resonance  after 10 years.  As can be seen in Figure~\ref{fig:3}, when both  effects, solar radiation force and  plasma drag, are present in the system, the particle leaves the resonance faster if only the solar radiation force is included. The semi-major axis starts increasing after the particles leaves the resonance due to the plasma drag. This representative particle goes to the outer edge of the E~ring.

Table~\ref{sitep} gives the lifetime of the CER sites due to the dissipative forces. The lifetime is defined as the time to $90\%$ of the initial sample of particles leave the CER or collide with the embedded moonlets, when moonlets are located in the system (section~\ref{immersed}). Since the particles located at the 7:6 CER are only under the effects of the solar radiation their lifetimes are larger compared to the lifetimes of the particles at the 14:15 and 10:11 CER. Since the particles at 10:11 CER are closer to Enceladus  (semi-major axis $\sim$ 238000~km \cite{Su17})  they have the smallest lifetime of the three systems  due to the  effects of the plasma drag.  The $1~\mu$m sized particles at the 14:15 and 10:11 CER last less than a decade.

\begin{table}
\centering
\caption{Time (in years) to 90\% of the  particles leave each arc.}
\label{sitep}
\begin{tabular}{cccccc} \hline\noalign{\smallskip}
     \multicolumn{2}{c}{$7$:$6$} & \multicolumn{2}{c}{$14$:$15$} & \multicolumn{2}{c}{$10$:$11$} \\
 $1~\mu$m  & $10~\mu$m  & $1~\mu$m   & $10~\mu$m   & $1~\mu$m   & $10~\mu$m  \\ \hline\noalign{\smallskip}
   150  &     $>$1000    &      10      &   80    &     5         &  34  \\ \hline\noalign{\smallskip}
\end{tabular}
\end{table}

\begin{figure}
\centering
\resizebox{1.05\columnwidth}{!}{\includegraphics{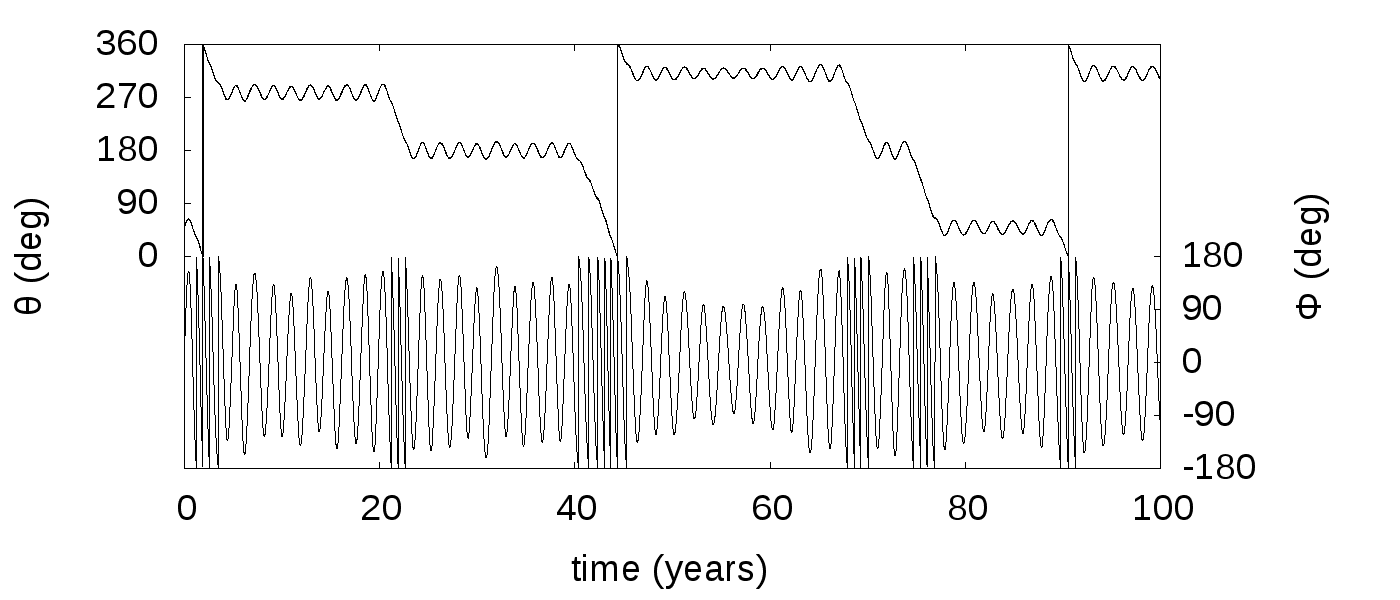}}
\caption{Time variation of $\theta$ and $\phi$. The $10~\mu$m sized particle is initially at $\Delta a=-40$ km and $\Delta\lambda=17^{\circ}$ from the centre of a $10$:$11$ CER site.}
\label{fig:4}       
\end{figure}

\begin{figure}
\centering
\resizebox{1.05\columnwidth}{!}{\includegraphics{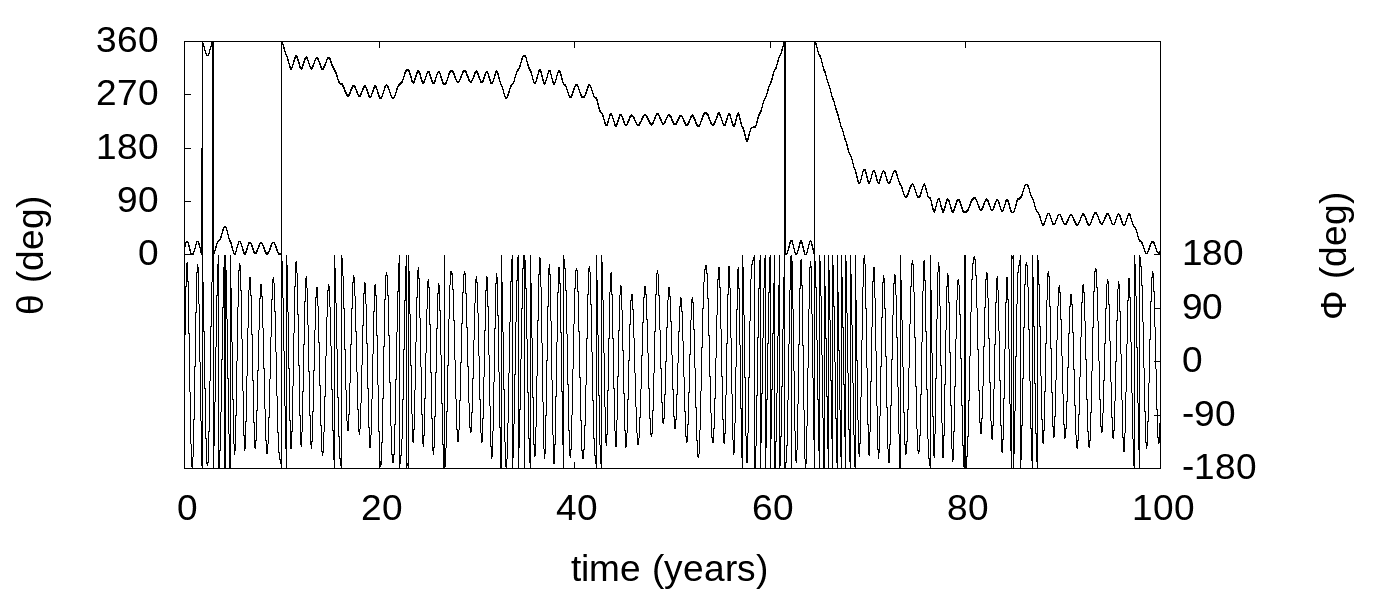}}
\caption{ Time variation of $\theta$ and $\phi$. The $10~\mu$m sized particle is initially at $\Delta a=70$ km and $\Delta\lambda=0^{\circ}$ from the centre of a $14$:$15$ CER site.}
\label{fig:41}       
\end{figure}

 Although the large majority of the particles are those represented in Figures~\ref{fig:2} and \ref{fig:3} a few of them may transit between the resonant arcs as shown in Figures~\ref{fig:4} and \ref{fig:41}. These figures present the time variation of the azimuthal angle ($\theta$) in the rotating reference  frame with the same velocity of the resonant frequency and the resonance angle ($\phi$) of a $10~\mu$m sized particle that leaves and returns to the resonance. In Figure~\ref{fig:4} the particle is initially at $\Delta a = -40$ km and $\Delta \lambda = 17^\circ$ from the 10:11 CER site and in Figure~\ref{fig:41} the particle is initially at $\Delta a = 70$ km and $\Delta \lambda = 0^\circ$ from the 14:15 CER site. As can be seen through the time variation of  $\theta$ and $\phi$, when the resonant angle $\phi$  librates the particles can be located in different values of $\theta$.  These are examples of particles travelling between the arcs.

\section{Immersed Moonlets} \label{immersed}

In this section are analysed the effects caused by an icy moonlet of different sizes located at the equilibrium point in the centre of the 7:6 CER site. The ring arc particles are under the effects of Saturn and its  gravity coefficients ($J_2$, $J_4$ and $J_6$), the satellite Mimas and a hypothetical embedded moonlet. Its  mass $m$  was assumed to be a fraction of  Mimas' mass ($\mu = m/M_M$), varying from $10^{-10}$  ($R\sim 100$~m) to $10^0$  ($R\sim 200$~km). The density of the moonlet is assumed to be  1~g/cm$^3$.

We numerically simulated the system using the Mercury integrator \cite{Ch99} with the Burlish St\"oer algorithm. The ring arc  is formed by a sample of 200 particles randomly chosen from the arc analysed in the last section. 

\begin{figure}
\centering
\subfigure[]{\resizebox{1.0\columnwidth}{!}{\includegraphics{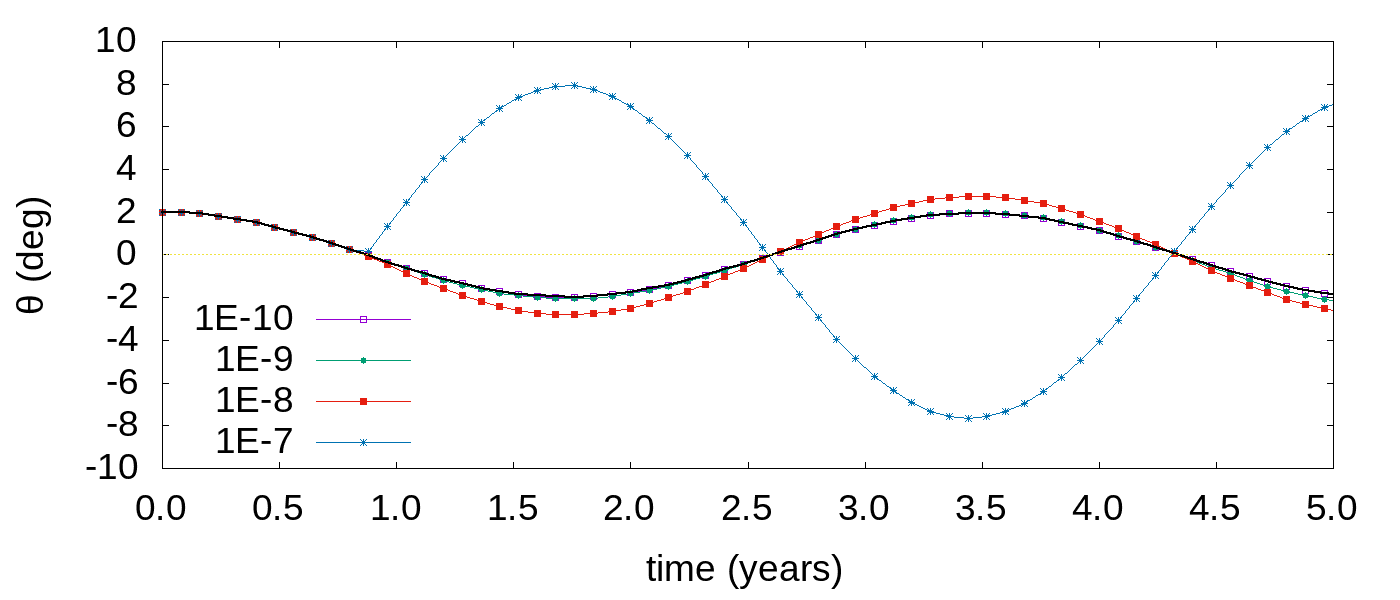}}}
\subfigure[]{\resizebox{1.0\columnwidth}{!}{\includegraphics{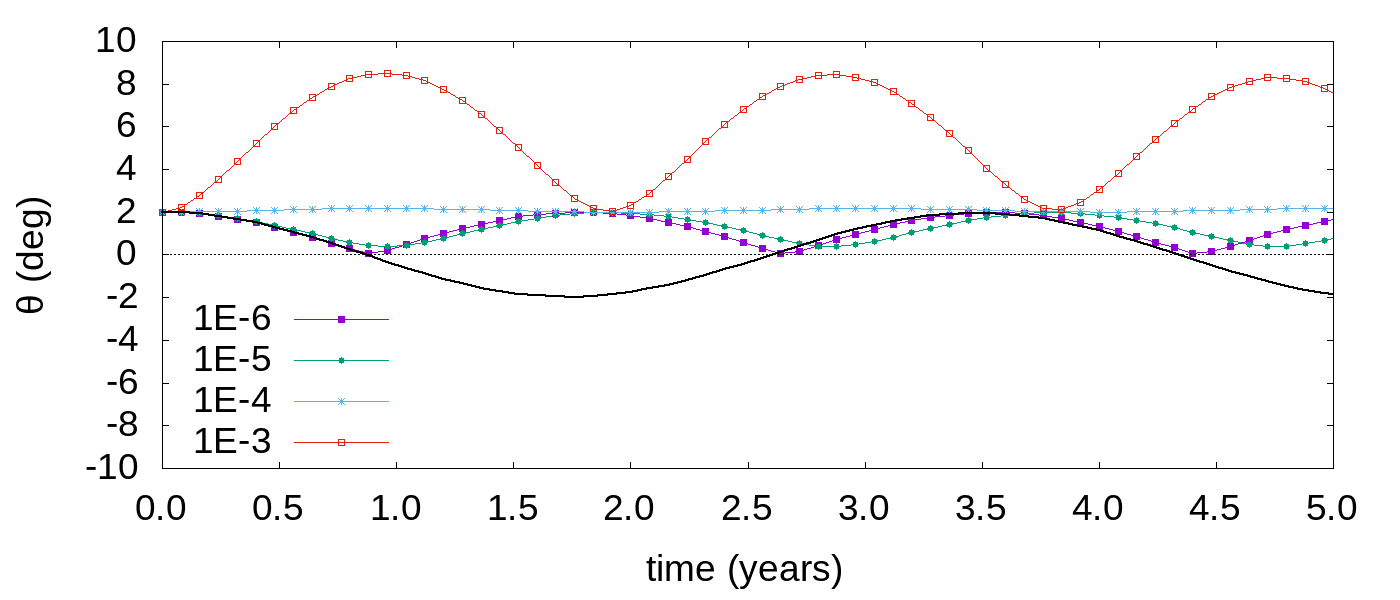}}}
\caption{Variation of  the azimuthal angle $\theta$ of a particle initially displaced $2^\circ$ from the centre of the corotation site for different values of $\mu$. The full line represents this variation when no immersed moonlet is located in the site. The presence of the moonlet can generate different regimes: a) first regime: for the mass ratios  of $10^{-10}$ and $10^{-9}$ the two curves overlapp, and b) second regime.}
\label{fig:5}       
\end{figure}

Figure~\ref{fig:5} shows the azimuthal angle $\theta$ of a particle initially located at $\Delta \lambda = 2^{\circ}$ from the centre of the 7:6 CER site. The moonlet is at the centre of the site. Different values of $\mu$ give rise to  three distinct regimes.  The black  thicker line shows the variation of $\theta$ when the moonlet is not in the system. Moonlets with $\mu = 10^{-10} -10^{-7}$, which we called the first regime,   provoke an increasing in the excursions of the ring arc particles  around the CER equilibrium point. In Figure~\ref{fig:5}a, which corresponds to the first regime, the azimuthal angle of the particle increases after a close approach with the moonlet  in less than one year. Larger moonlets induce larger variations in $\theta$.

\begin{figure}
\centering
\subfigure[]{\resizebox{1.0\columnwidth}{!}{\includegraphics{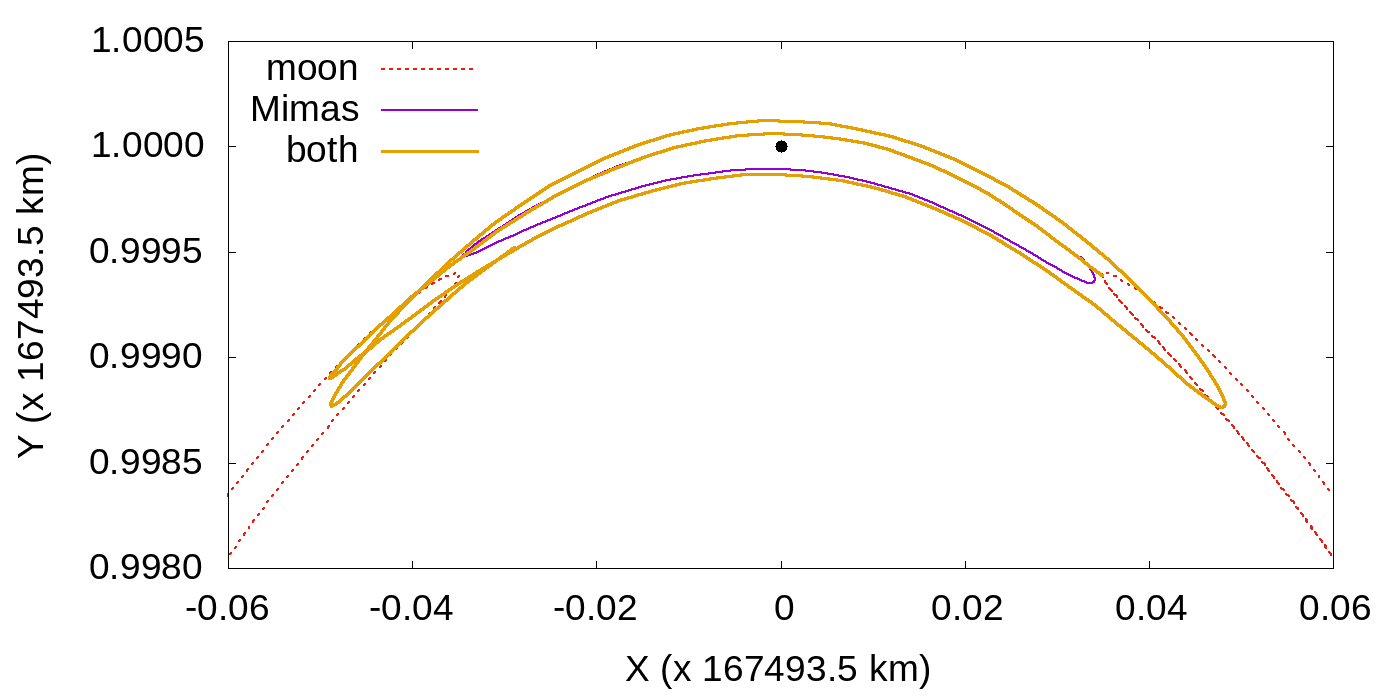}}}
\subfigure[]{\resizebox{1.0\columnwidth}{!}{\includegraphics{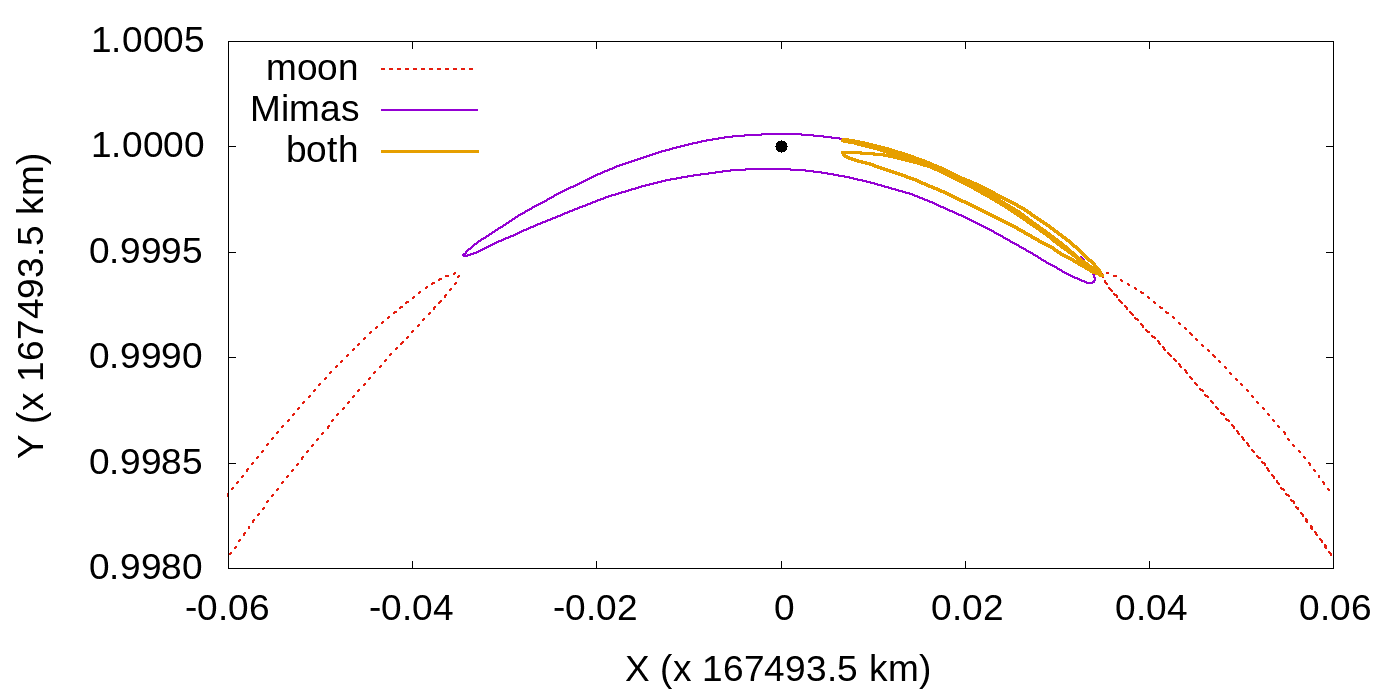}}}
\subfigure[]{\resizebox{1.0\columnwidth}{!}{\includegraphics{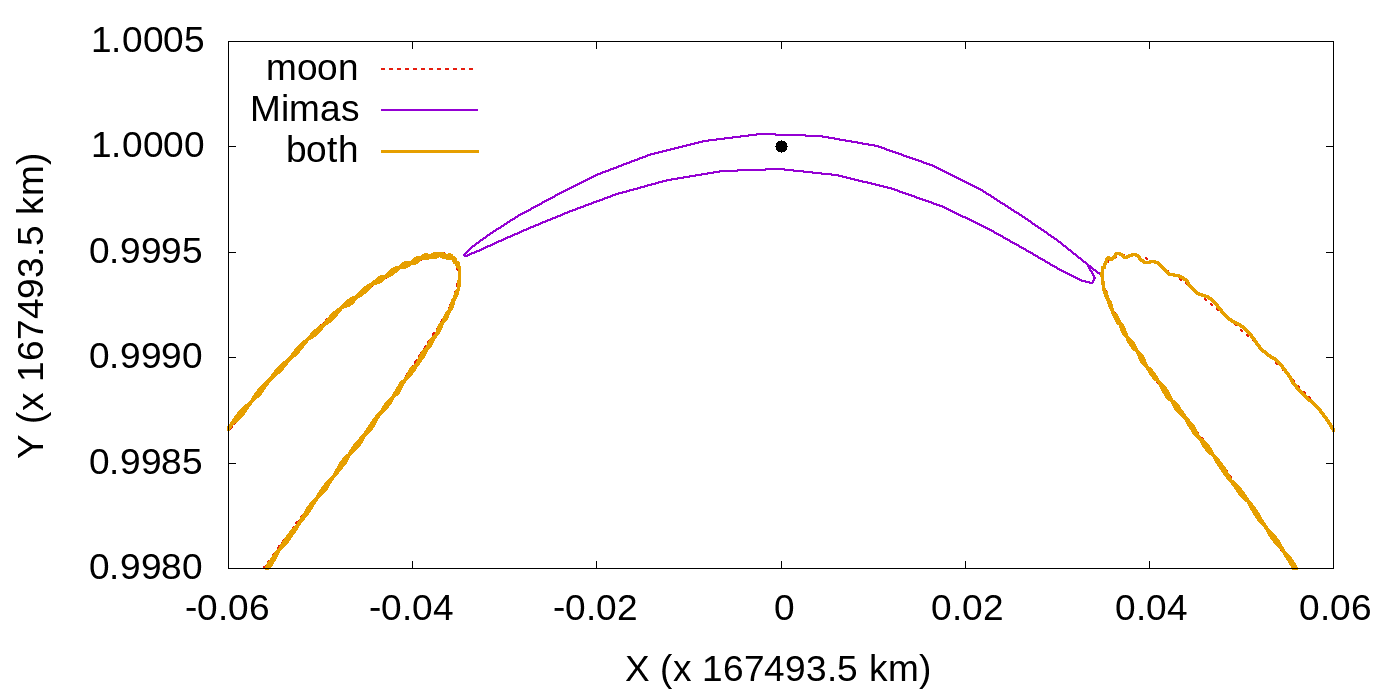}}}
\caption{Motion of particles initially placed $2^\circ$ from the centre of the 7:6 CER site for a) $\mu = 10^{-8}$, b) $\mu = 10^{-5}$ and c) $\mu = 10^{-2}$.  When the system is formed by Saturn and Mimas  the motion of the particle is represented by a full line (arc), when the system is formed by Saturn and the moonlet its motion is shown in dashed line (horseshoe orbit) and when both satellites are present (Mimas and the moonlet) the motion of the particle can be seen in a thicker full line. }
\label{fig:6}       
\end{figure}

Moonlets ranging from $10^{-6}$ to $10^{-3}$,  second regime,  are responsible for azimuthally confining  the particles. 
 These particles stay confined in half part of the arc. In Figure~\ref{fig:5}b, second regime, the perturbation of the moonlet provokes a change in the motion of the particles. The particles  did not move around the CER equilibrium point but remain located between the moonlet and the edge of the arc. In the third regime which corresponds to a moonlet with $\mu \geq 10^{-3}$  the perturbation of the moonlet removes the particle from the resonance with Mimas.  The particle stays in a horseshoe orbit with the moonlet (see figure~\ref{fig:6}c).  

The motion of the particles in the rotating reference system can be seen in Figure~\ref{fig:6}. Three values of $\mu$ are shown: a) $\mu = 10^{-8}$  corresponding to a moonlet with radius of 500~m, b) $\mu = 10^{-5}$  corresponding to a moonlet with radius of 5~km and c) $\mu = 10^{-2}$  corresponding to a moonlet with radius of 50~km. The dot shows the position of the hypothetical moonlet. The motion of the particle in the system formed only by Saturn and Mimas is shown in full line (the smallest arc), when  only Saturn and the moonlet are  in the system the behaviour of the particle is represented by a dashed line (a horseshoe orbit) and when both perturbers are present, Mimas and the moonlet, the motion of the particle can be seen in a thicker full line.

In Figure~\ref{fig:6}a the particle starts its motion under the gravitational effects of Mimas. However due to the moonlet the particle completes an arc larger than if it was only under the effects of Mimas. In Figure~\ref{fig:6}b as the particle approaches the moonlet it returns to a small arc becoming confined in half part of the corotation site. In Figure~\ref{fig:6}c the particle is removed from the site due to the effects of the large moonlet and becomes confined in a horseshoe orbit. The perturbation of Mimas causes only small variations in the horseshoe orbit.
 
Similar results were found for the 14:15 and 10:11 CER sites when an immersed moonlet is located in its centre.

\section{Fate of the CER Sites} \label{fate}

Immersed moonlets on planetary arcs or rings can help to replenish those particles lost by collisions or ejection from the system.  Interplanetary dust particles  (IDPs)  can collide with these moonlets; these collisions  produce  material which can be ejected from their surfaces. These IDPs, originate  from the Jupiter family and the Kuiper Belt,  travel through the interplanetary environment  with velocities of order of kilometres per second \cite{Po16}.

Following the algorithm presented by \cite{Sf12}, the mass production rate ($M^+$)  given by a moonlet with radius $R$ is 
\begin{equation}
M^+=\pi R^2F_{imp}Y    
\end{equation}
 where $F_{\rm imp}$ is the mass flux of impactors and $Y$ is the ejecta yield.
 
 Reference \cite{Po16} estimates that the impactors mass flux at Saturn's region is of order $F_{imp}^{\infty} = 10^{-17}$~kg/(m$^2$s). The mean velocity of the impactors is $v_ { imp}^{\infty} = 9.5$~km/s \cite{Sp06B}. As the impactors approach the planet, the mass flux and also the velocity are enhanced due to the gravitational focusing of the planet. At  distance $a$ from Saturn's centre, the impactor mass flux  $F_{ imp}$ and $v_ { imp}$ are \cite{Kr03}

\begin{equation}
\frac{\upsilon_{imp}}{\upsilon_{imp}^{\infty}}=\sqrt{1+\frac{2GM_S}{a(\upsilon_{imp}^{\infty})^2}}
\end{equation}
\begin{equation}
\frac{F_{imp}}{F_{imp}^{\infty}}=\frac{1}{2}\left(\frac{\upsilon_{imp}}{\upsilon_{imp}^{\infty}}\right)^2+\frac{1}{2}\frac{\upsilon_{imp}}{\upsilon_{imp}^{\infty}}\left[\left(\frac{\upsilon_{imp}}{\upsilon_{imp}^{\infty}}\right)^2 
-\left(\frac{R_S}{a}\right)^2\left(1+\frac{2GM_S}{R_S(\upsilon_{imp}^{\infty})^2}\right)\right]^{1/2}
\end{equation}

The ejecta yield depends on the composition of the moonlet. \cite{Ko01} found that the ejecta yield for icy satellites hit by $\mu$m-sized projectiles is
\begin{equation}
Y=2.64\times10^{-5}m_{imp}^{0.23}~\upsilon_{imp}^{2.46}    
\end{equation}
where $m_{imp}$  is the mass of the impactor. This value was assumed to be $10^{-18}$~kg in Saturn's region \cite{Sp06B}.

The value of the mass production rate enable us to estimate the time $t_p$ necessary to the moonlet populates the CER site with a sample of $\mu$m-sized dust particles (1-10~$\mu$m):
\begin{equation}
t_p=\frac{m_s}{M^+} \label{tp}
\end{equation}
where $m_s$ is the mass of the site given by \cite{Sf12}:
\begin{equation}
m_s=A_s\left(\frac{4}{3}\pi\rho\right)\int_{1\mu {\rm m}}^{10\mu {\rm m}}\pi r^3dN
\end{equation}
$\rho$ is the density of the particle ($\rho=1$~g/cm$^3$) and $A_s$ is the surface area of the site.

We assumed the distribution of dust particles to be $dN = C~r^{-3.5}~dr$ \cite{Co93}, where $C$ is a constant determined by the optical depth \cite{Sf12}
\begin{equation}
\tau=\int_{1\mu {\rm m}}^{10\mu {\rm m}}\pi r^2dN    
\end{equation}

The optical depth of the G~ring arc was assumed to be $\tau = 10^{-5}$,  and for the arcs of Methone and Anthe the adopted value is $\tau = 10^{-7}$ \cite{He09,He10}. 

To determine $A_s$  we assumed the site as a ring segment with angular width of $360^{\circ}/m$ and internal and external edges as $(a_c - W_c/2)$ and $(a_c + W_c/2)$ from the centre of the planet, respectively.  $A_s$ can be written as
\begin{equation}
A_s=\frac{2\pi}{m}a_c~W_c \label{As}
\end{equation}

The algorithm described above was used to calculate  how long  the moonlet takes to populate the site. 

We also computed the lifetime of the CER site particles through a sample of numerical simulations for a time span of 1000~years. Each site (7:6, 14:15 and 10:11  CER) was composed by a set of particles and an immersed moonlet located in its centre,  under the perturbations of Mimas and the gravity coefficients of Saturn. The size of the moonlet  was assumed to be proportional  to the nominal satellite located in each arc of the Saturn system. Table~\ref{labels} shows the values of the  radius of the moonlets.  Moonlets larger than the 2.5z case have ejection velocities greater than the average velocity of the dust produced by IDPs processes and the equation~\ref{tp} cannot be applied.  We performed numerical simulations without and with both dissipative forces, solar radiation force and plasma drag. The arcs are formed by particles with  1$\mu$m and $10~\mu$m in radius. 

\begin{table}[]
\centering
\caption{Radius of the moonlets immersed in the centre of each site and its label.}
\label{labels}
\begin{tabular}{rccc} \hline \hline
\multicolumn{2}{c}{moonlet}              & \multicolumn{1}{l}{radius (km)} & \multicolumn{1}{l}{label} \\ \hline
50 \%  & \multirow{3}{*}{radius of Aegaeon} & 0.13                          & 0.5x                       \\
100 \% &                                  & 0.25                          & 1.0x                       \\
150 \% &                                  & 0.38                          & 1.5x                        \\ \hline
100 \%  & \multirow{2}{*}{radius of Anthe} & 0.50                          & 1.0y                       \\
150 \% &                                  & 0.75                          & 1.5y                       \\ \hline
100 \%  & \multirow{4}{*}{radius of Methone} & 1.40                          & 1.0z                       \\
150 \% &                                  & 2.18                          & 1.5z                      \\ 
200 \%  &                                 & 2.90                          & 2.0z                       \\
250 \% &                                  & 3.63                          & 2.5z                       \\ \hline
\end{tabular}
\end{table}

\begin{figure}
\centering
\subfigure[]{\resizebox{0.7\columnwidth}{!}{\includegraphics{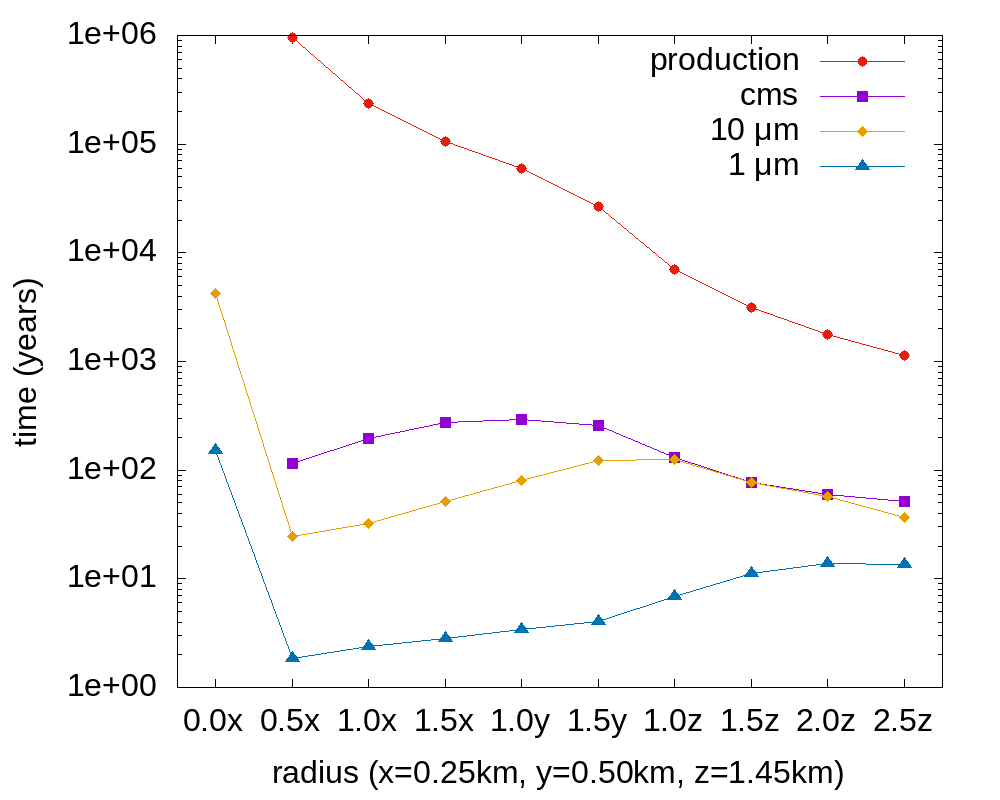}}}
\subfigure[]{\resizebox{0.7\columnwidth}{!}{\includegraphics{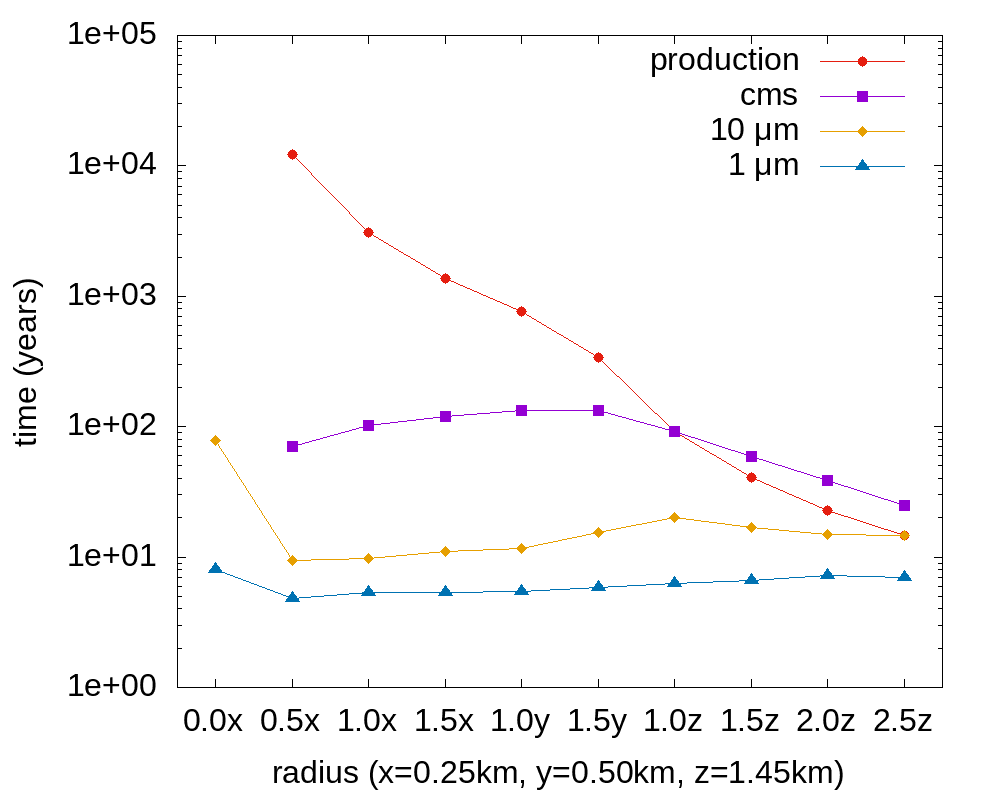}}}
\subfigure[]{\resizebox{0.7\columnwidth}{!}{\includegraphics{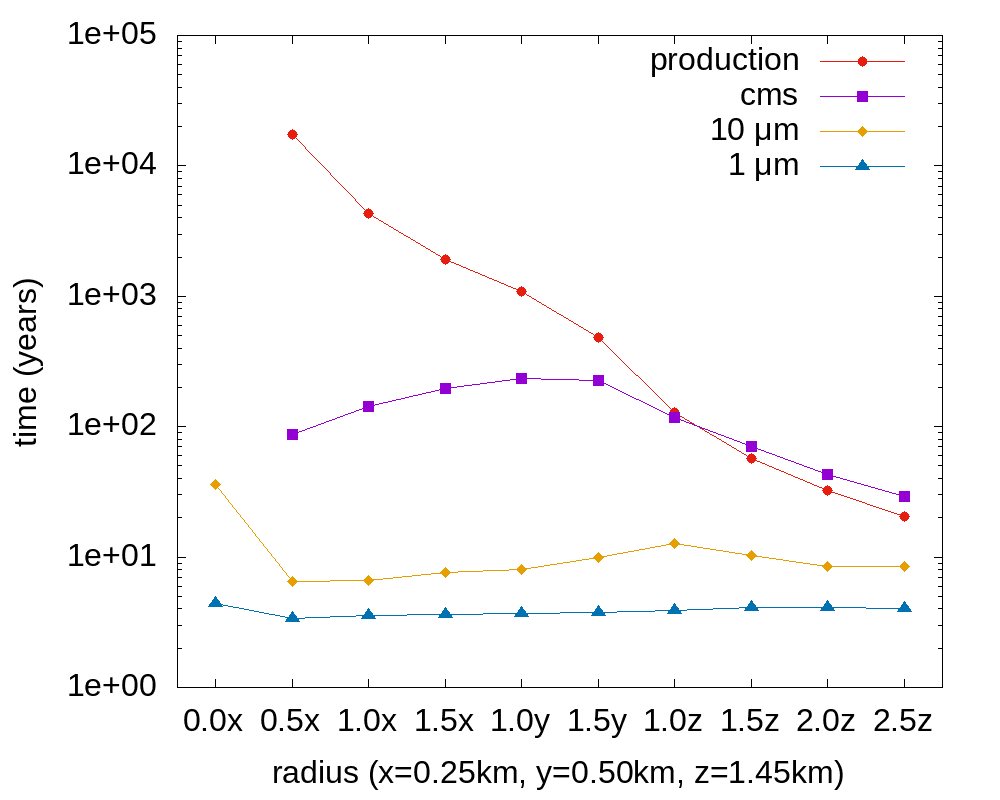}}}
\caption{Time (in years) versus size of the moonlet. This figures shows the time for the moonlet produces material (``production'') and removes it for two cases, without {\bf (``cms'')} and with   dissipative forces. For the last case two samples of particles were analysed: $1~\mu$m (``$1~\mu$m'') and $10~\mu$m (``$10~\mu$m'') sized particles. Each plot represents each site:  a) 7:6~CER, b)  14:15~CER and c)  10:11~CER. }
\label{fig:7}       
\end{figure}

Figure~\ref{fig:7} shows the time (in years) versus  different sizes of a moonlet  immersed in the arc. The curve which shows the time that each moonlet needs to populate the arc  with a sample of 1-10~$\mu$m dust particles (derived from equation~\ref{tp}) was labelled  as ``production".
The time for 90\% of all the particles leave the site or collide with the moonlet without any dissipative force was labeled as ``cms''. In this case, the arc particles are only under the gravitational perturbation of Mimas and the moonlet and can represent the motion of cm-sized particles if they are in the arcs. For this size range  the effects of the dissipative forces can be neglected.

When  dissipative forces, solar radiation force and plasma drag, are present in the system, there are two curves labelled as:  ``$1~\mu$m"  and ``$10~\mu$m". The first one represents the arc populated by  $1~\mu$m sized particles while the curve labelled ``$10~\mu$m'' refers to the arc populated by $10~\mu$m sized particles. Note that these two curves have values for 0.0$\times$ (in the $x$ axis), which means  that the arc has none immersed moonlet and its lifetime is derived from  Table~\ref{sitep}.

Figure~\ref{fig:7}a represents the  7:6~CER simulated arc.  The ``production'' curve decreases as the size of the moonlet increases,  larger moonlets need less time to populate the arc, as expected. The lifetime of the arc particles, when no dissipative force is present in the system,  can be seen in the {\bf ``cms''} curve which shows the first and second regimes, as discussed in section~\ref{immersed}. The first regime corresponds to moonlets which sizes from $0.5 x$ (radius = 0.13~km) to $1.5 y$ (radius = 0.75~km), the  ``cms'' curve increases up to the  size $1.5 y$. In this regime the particles  go further from the moonlet. For larger moonlets ($> 1.5y$) the particles are confined in the second regime and closer to the moonlet,  leading the particles to collisions with them. Both ``$1~\mu$m"  and ``$10~\mu$m" curves represent the lifetime of the particles when  the solar radiation force is acting in the system. The ``$10~\mu$m" curve is very similar to the  ``cms'' curve, since the solar radiation perturbation causes only small increase in the eccentricity of the $10~\mu$m sized particles. The only difference is that the size of the  moonlet  separating  the first to the second regime is $1.0z$ (radius = 1.40~km), larger than for the  ``cms'' curve. Accordingly to Figure~\ref{fig:7}a the time for the moonlets  produce arc particles is always larger than the lifetime of the particles. Therefore the 7:6~CER arc is a transient arc. 

Figure~\ref{fig:7}b represents the 14:15~CER arc. Both curves, the ``production'' and the {\bf ``cms''} curves, present almost  the same behavior as discussed in Figure~\ref{fig:7}a. There is one difference,  for larger moonlets ($> 1.0z$) the time to produce  dust particles to supply those  lost by the arc is smaller than the lifetime of the arc.  This arc could survive if larger particles populate it. However, another process would be need to replenish the arc since the IDPs process is able to provide only micrometer dust particles to the arc. The ``$10~\mu$m" curve presents similar behavior as analysed in Figure~\ref{fig:7}a, however in this case  (Figure~\ref{fig:7}b) both dissipative forces are present. The ``$1~\mu$m" curve seems to be independent of the size of the moonlet, the lifetime of the arc populated by $1~\mu$m sized  particles is about 10 years regardless the size of the immersed moonlet.

Figure~\ref{fig:7}c represents the 10:11~CER arc. The analysis  of the behavior of the curves  in Figure~\ref{fig:7}b is the same for Figure~\ref{fig:7}c. The arcs of Anthe and Methone have similar dynamical behavior.

\section{Discussion} \label{discussion}
The arcs in Saturn system have immersed satellites on them. The G~ring arc has the small satellite Aegaeon, while  the arcs of Anthe and Methone  have both satellites Anthe and Methone present.
These arc particles are also under the resonant effects of the satellite Mimas. Aegaeon, Anthe and Methone
are trapped in a 7:6, 10:11 and 14:15 CER with Mimas. The  population of these arcs is mostly formed by $\mu$m-sized particles, which are disturbed by dissipative forces such as the solar radiation force.

In this work we analysed the lifetime of these arcs  under the gravitational effects of Saturn, Mimas, the small satellites and dissipative forces. For the arcs of Methone and Anthe, since they are closer to Enceladus, the effects of the plasma drag have to be taken into account. As a result we found that the $1~\mu$m sized particles live longer in the G~ring arc (about 150 years) while in the arcs of Anthe and Methone they can last at most one decade.  Larger particles ($10~\mu$m in radius) located in the G~ring arc, Methone and Anthe arcs  have lifetimes larger than 1000 years, about 80 and 35 years, respectively.

An immersed moonlet can affect the arc population in two ways. Firstly, the moonlet can supply the arc population with  particles  released from its surface  after a collision with interplanetary objects.  The moonlet can be the source of the arc. By the other hand, the moonlet can disturb the particles and provokes collisions and ejections from the arc. The moonlet can also be the sink of the arc.
Therefore we analysed a  hypothetical scenario by assuming  a  moonlet of different sizes present in each arc.  We calculated the mass production  of each moonlet and the orbital evolution of the particles, and consequently their lifetimes, under the effects  of the moonlet.

Our results, after a sample of numerical simulations,  have shown that a set of particles under the gravitational effects of Mimas and a hypothetical moonlet  can perform three different motions. When the moonlet has mass ratio between $10^{-10}$ to $10^{-7}$ the particle stays in an arc   larger than if only Mimas was present in the system. As the moonlet mass increases up to $10^{-3}$, the moonlet reduces the motion of the particle. It stays in half part of the arc. For moonlets  larger than $10^{-3}$ the particle performs a horseshoe orbit  disturbed by Mimas.

 By comparing the lifetime of the arcs (without and with the dissipative forces) and the mass production due to the moonlets we found that all three arcs are transient. The dissipative forces dominate the micrometric particles dynamics, regardless of their size. They survive in the arc up to a few hundred of years while the time for a moonlet replenish the arc is always longer than the particles lifetime, according to our dynamical model. A cm-sized population could survive in the arc for a longer period of time however these particles cannot be supplied by the IDPs process. In general, we found that the arcs are rapidly depleted of micrometric dust particles and they tend to be transient structures.

\section*{Acknowledgements}

The authors thank  Fapesp (2016/24488-0, 2016/24561-0 and 2018/23568-6) and  CNPq (309714/2016-8) for the financial support. This study was financed in part by the Coordena\c c\~ao de Aperfei\c coamento de Pessoal de N\'\i vel Superior - Brasil (CAPES) - Finance Code 001. 

GM and SMGW divided the work of writing and the numerical simulations.


\end{document}